\documentclass[aps,amsmath,twocolumn]{revtex4}
\usepackage{graphicx}
\begin{document}

\title{Avoidance of instability of a superluminal Gaussian light pulse via control of nonlinear coherence Kerr effect in a gain-assisted medium
}
\author{Bakht Amin Bacha}
\affiliation{Department of Physics, Hazara University, Pakistan}
\author{Fazal Ghafoor}
\affiliation{Department of Physics, COMSATS Institute of
Information Technology, Islamabad, Pakistan}

\begin{abstract}
We investigate nonlinear Kerr-induced coherence effect on a
superluminal probing light pulse in a gain-assisted N-type 4-level
atomic system via an intense monochromatic laser field. The
dispersion exhibits a novel, interesting and useful two-paired
double gain lines processes. The system displays lossless
characteristics similar to [L. J. Wang, A. Kuzmich, A. Dogariu,
Nature \textbf{406}, 277 (2000)] but with advantages of
\emph{multiple} \textbf{controllable} anomalous regions,
\emph{significantly enhanced} superluminal behavior and
\textbf{relaxed} temperature, states of matter regardless of its
isotropic or anisotropic conditions. Unlike the instability in [A.
M. Akulshin, S. Barreiro, and A. Lezama, Phys. Rev. Lett.
\textbf{82}, 4277 (1999)], the present system also \emph{overcomes
the instable-limit} by the Kerr-induced coherence effect in the
system. Indeed, the coherence enhances the group velocity
remarkably by at least $-30 ms$ more than of an instable Kerr-free
system with almost negligible distortion in the retrieved pulse.
\end{abstract}

\maketitle
The significant development of theoretical and
experimental techniques for the control of light pulse propagation
through resonance optical media had taken place in the last few
decades. The control over the properties of spontaneous emission
\cite{Zhu-Scully,Paspalakis,darkatom,2-atom}, stimulated
absorption \cite{Absorption}, dispersion\cite {Absorption01} and
group index of media \cite{Index} had led to the observation of
some fascinating phenomena such as coherent population trapping,
lasing without inversion \cite{EA96,AS9596}. In most of these
types studies, coherent fields were used to control the optical
properties of the media \cite{sg04}. In fact, their practical and
theocratical aspects were also investigated for slow and even for
stopped (stored) light pulse \cite{AK95,OS96,MM99,ms03,LS01}.

The superluminal light pulse propagation had also been remained
under extensive studies among researchers. One of these studies
was reported by Bigelow \emph{et al.} via beating of a pump and
probe field to generate coherent population pulsation in the
medium \cite{ms03}. Similarly, a negative group velocity was
studied in a medium with a gain doublet \cite{AM1994}, and in a
transparency medium with inverted population \cite{RY1982}. The
list of such interesting studies is quite long, however some of
these can be found in Refs.
\cite{Nature,physd,review,FCARRELIO2005,AM1999,KKIM2003,HKANG2003,FXIAO2004,tj04}.
Be specific, we concentrate on two pioneer and sophisticated
experiments, and their complications. In one
\cite{Wang2000,Wangprl,Wangpra,Cao}, Wang \textit{et al.} observed
superluminal behavior of a probing light pulse using the region of
lossless anomalous dispersion between two closely spaced gain
lines. However, due to limited capability of the experiment, the
light velocity could not be increased more than c/-310 (where c is
the velocity of light in vacuum). In the other remarkable
experiment, Akulshin and his co-workers defined a stability limit
for the superluminal probing pulse constrained by the strength of
the bi-chromatic laser field \cite{Akulshin01}. Recently, the
incompatibility of a pulse width with a steep anomalous region
created a pulse distortion and the other related problems [see for
example
\cite{PRAoptimal,eurolett,opticslett,opticslett01,PRA-instable}].
Consequently, the applied aspects of these studies were limited.
For example, precession temporal cloaking \cite{Moti2012},
precession image measurement \cite{RYANT2012} and may be many
other, were suffered from a limited group velocity of the pulse
when media of negative group index were used \cite{Robret2012}.
Therefore, enhancing superluminality while avoiding the earlier
studied complication, in general, is an important task.

The Kerr effect is a nonlinear and noise free phenomenon used by
Hai and co-workers for nonlinear index of a three-level atomic
medium inside an optical ring cavity. The greatly enhanced Kerr
effect near atomic resonance condition led them to measure group
index up to $7.0 ×10^{-6}cm/w$ \cite{hwanggroosky2001}. The
enhanced nonlinear kerr effect was also investigated both
theoretically and experimentally in different systems in different
contexts \cite{hwanggroosky2001,hchang2004,hkang2003}. The various
aspects of the Kerr effect were used for quantum phase gate,
optical solitons \cite{ov03,ov06,wd04,HH06}, quantum logic gate,
quantum non-demolition, quantum teleportation, and nonlinear light
control \cite{ jp1993,QA1995,HW2002}. Dey and Agarwal used the
coherence generated by the Kerr effect to slow-down light pulse
through an Electromagnetically Induced Transparency (EIT) medium
\cite{TNGS2007}.

In this letter, we investigate Kerr-induced coherence effect on a
superluminal weak probing pulse in a gain-assisted medium via a
strong monochromatic laser field. Intriguingly, the coherence
created by the Kerr effect in the proposed atomic system is
significant for the propagating pulse. Consequently, the generated
coherence overcomes the instable limit of our system unlike the
system of the Ref. \cite{Akulshin01}, where the limit was created
by the incoherence effect of the strengthened intensity (power
broadening) of the bi-chromatic pump field. Nevertheless, the
group index of our system gets a direct proportionality with the
square of the Kerr-field intensity caused by the Kerr-induced
coherence effect. Consequently, the superluminality is enhanced
even with a less losses (undistorted retrieved pulse) like those
of Ref. \cite{Wang2000} but with the advantages of multiple
controllable anomalous regions, relaxed temperature, states of
matter, and their isotropy or anisotropy conditions. Obviously,
undistorted retrieved pulse is normally necessary for practical
applications and the present system exhibits the same
characteristics when analyzed analytically.
\begin{figure}[t]
\centering
\includegraphics[width=2.8in]{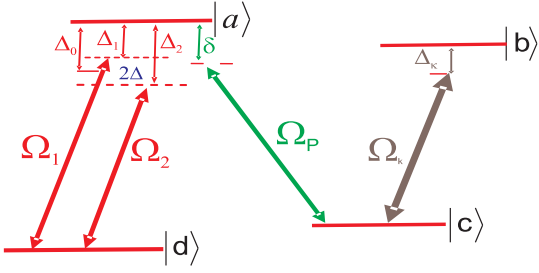}
\caption{ Schematics of the proposed system }\label{figure1}
\end{figure}

Three-level atomics Cesium of a gaseous medium in a vapor cell
were considered by Wang \emph{et al.} in their experimental
set-up. Two strong continues-wave Raman pump light beams $E_1$,
and $E_2$, of frequencies $\nu_1$, and $\nu_2$ (with difference
$2\Delta=\nu_2-\nu_1$ ), were allowed to interact with the Cesium
atoms kept at temperature $30^\circ C$. The upper excited energy
level was considered as $6 P_{3/2}$ $\left|F=4,m=4\right\rangle$
$\left|a\right\rangle$ while $6 S_{1/2}$
$\left|F=4,m=-4\right\rangle$ $\left|d\right\rangle$ and $6
S_{1/2}$ $\left|F=4,m=-2\right\rangle$ $\left|c\right\rangle$ were
assumed as the two ground states of the atom. Also, a probe field
is coupled with the $\left|c\right\rangle$ and
$\left|a\right\rangle$. Both the fields are detuned from the
transition frequency $\nu_{ad}$, [$\left|a\right\rangle
to\left|d\right\rangle$], by a large average detuning $\Delta_0$.
The optical susceptibility of the probe field was calculated as:
\begin{eqnarray}
\chi_w=[\frac{M_1}{\nu_p-\nu_1+i\gamma}+\frac{M_2}{\nu_p-\nu_2+i\gamma}],
\end{eqnarray}
where
$M_{1,2}=N|\sigma_{ac}|^2|\Omega_{1,2}|^2/4\pi\epsilon_0\hbar\Delta^2_{0}$.
Using experimental data they obtained the group index $-310\pm5$,
corresponding to the advanced time $-62 ns$.

We consider an N-type four-level atomic system driven by two pump
fields and a probe field. Meanwhile, we couple a monochromatic
laser field, which we call the Kerr field, due to its induced
coherence effect on the dynamics of the system [see Fig. 1]. The
two lower levels $\left|d\right\rangle$ and $\left|c\right\rangle$
are coupled with the upper level $\left|a\right\rangle$ by two
coherent pump fields and a weak probed field of Rabi frequencies $
\Omega_1$, $\Omega_2$, and $\Omega_p$, respectively. Furthermore,
the coupling of the intense Kerr field is with the levels $
\left|c\right\rangle$ and $\left|b\right\rangle$ having the Rabi
frequency $\Omega_k$. Now to present the model and equations of
motion, we proceed with the following interaction picture
Hamiltonian in the dipole and rotating wave approximations:
\begin{eqnarray}
H_I&=&-\frac{\hbar}{2}( \Omega_1 e^{-i\Delta_1t}+ \Omega_2
e^{-i\Delta_2t})\left|a\right\rangle\left\langle d\right|  \nonumber \\
&& -\frac{\hbar}{2}\Omega_k
e^{-i\Delta_kt}\left|b\right\rangle\left\langle
c\right|-\frac{\hbar}{2}\Omega_p e^{-i\delta
t}\left|a\right\rangle\left\langle c\right|+H.c,
\end{eqnarray}
where $\nu_1=\nu_{ad}\pm\Delta_1$, $\nu_2=\nu_{ad}\pm\Delta_2$, $%
\nu_k=\nu_{bc}\pm\Delta_k$ and $\nu_p=\nu_{ac}\pm\delta$, while
$\nu_{1}$, $\nu_{2}$, $\nu_p$, $\nu_{k}$, are the frequencies of
the two pump fields, the probe field and the Kerr field,
respectively. The detuning parameters appearing in the system are
$\Delta_1=\Delta_0-\Delta$, and $\Delta_2=\Delta_0+\Delta$, where
$\Delta=(\nu_2-\nu_1)/2$ is the effective detuning,
$\Delta_0=(\Delta_1+\Delta_2)/2$ is the average detuning. We use
the following general form of density matrix equation:
\begin{equation}
\frac{\partial\rho }{\partial t}=-\frac{i}{\hbar
}[\rho,H_{I}]+\Lambda \rho
\end{equation}%
for the equations of motion, where $\Lambda \rho $ is the damping
part of our system. The equations of motion for slowing varying
amplitudes $\widetilde{\rho}$ for different matrix elements are
then evaluated using its transformation relation with the fast
varying amplitudes We evaluated the steady state solution for
$\widetilde{\rho}_{ac}$ and obtained
$\widetilde{\rho}_{ac}=\frac{-i\Omega_p(\beta_1+\beta_2)}{8}$. The
parameters $\beta_1$, and $\beta_2$, are listed in the appendix-B.
The susceptibility, which is a response function to the applied
fields, is obtained as
\begin{eqnarray}
\chi=\frac{2N|\sigma_{ac}|^2\widetilde{\rho}_{ac}}{\epsilon_0\hbar\Omega_p}=\frac{-i3N
\gamma\lambda^3}{32\pi^3}(\beta_1+\beta_2),
\end{eqnarray}
where $N$, is the atomic number density, $\lambda=2\pi
c/\omega_{ac}$, is the wavelength, and
$A=4\sigma_{ac}|^2\omega_{ac}^3/\epsilon_0\hbar c^3=4\gamma$, is
the Einstein coefficient. Consistently, our main results is
reducible to the theoretical result of Wang \emph{et al} for the
dispersion and absorption of the probe light pulse. In this
connection, if we assume $\gamma_{ad}=\gamma_{bd}=0$,
$\Omega_k=0$, $\Delta_k=0$, $\gamma_{ac}=\gamma_{cd}=\gamma$, and
$\Delta_0\approx\delta$ while
$\Delta/\Delta_0,\gamma/\Delta_0<<1$, we obtained
$\chi=M_1/(\delta-\Delta_1+i\gamma)+M_2/(\delta-\Delta_2+i\gamma)$,
where $M_{1,2} =
N|\sigma_{ac}|^2|\Omega_{1,2}|^2/4\epsilon_0\hbar\Delta^2_{0}$,
[$\sigma_{ac}$, appears for the corresponding dipole moment].
Further, if $\nu_{ac}\approx\nu_{ad}$, and
$\Delta_{1,2}=\nu_{1,2}-\nu_{ad}$, while
$\delta=\nu_{p}-\nu_{ac}$, the results obtained is similar to Eq.
(1) as calculated by Wang \emph{et al.}

Generally, the Kerr effect means a change in the refractive index
of a material in response to an applied electric field. It is the
electric field intensity dependent nonlinear effect on the
response function of the medium. The corresponding refractive
index is proportional to the square or higher order power of the
field. Therefore, the change in the refractive index can be
measured from $\Delta n_r=\lambda K E^{2,4,6,8..}_0,$ where
$\lambda$ is the wavelength and $E_0$ is the amplitude of the
electric field, while K is the Kerr coefficient. Specifically, the
Kerr effect in a medium appears due to intense laser light. If
un-important higher order terms are neglected, then the Kerr
effect is given by $n_r=n_0+n_2I$, where $n_r=\sqrt{1+\chi}$
\cite{pb2012,ea2011}. One of our main task is to explore a
mechanism in the system for an enhancement of superluminality. The
intense Kerr field may achieve this task under favorable
conditions for superluminal light. These favorable conditions
includes: both low (high) temperatures regimes for homogenous
(nonhomogeneous) gaseous, liquid, and solid state media.
Nevertheless, unfavorable conditions are very large optical
bandwidth and the saturation limit of the Kerr effect

Now, a formalism for the contribution from the Kerr effect is
developed by the expansion of $\chi$ in the power series of the
Rabi frequency $\Omega_k$, if the unimportant higher order terms
are neglected \cite{TNGS2007}. Therefore we write
\begin{equation}
\chi^k=\chi^{(0)}+\frac{1}{2}|\Omega_k|^2 \frac{\partial^2\chi
}{\partial\Omega^2_k}|_{\Omega_k\rightarrow 0.}
 \end{equation}
Here, the first term $\chi^{(0)}$ represents the susceptibility
when the Kerr field is zero while the second term represents
susceptibility contributed by $|\Omega_k|^2$. Obviously,
$|\Omega_k|$ is the atom-Kerr-field coupling constant which
depends on the Kerr field intensity. The greater is the intensity
of the field, the larger is the Kerr effect in the system. To
explore the induced coherence in the system by the Kerr effect we
estimated the susceptibility as
\begin{equation}
\chi^k=\frac{-3iN
\gamma\lambda^3}{32\pi^3}[\beta_3+T_1+T_2+\frac{1}{2}|\Omega_k|^2(
K_1+K_2)],
\end{equation}
where $ \chi^{(0)}=\frac{-3iN
\gamma\lambda^3}{32\pi^3}[\beta_3+T_1+T_2]$. In this system we
dealt with the third order susceptibility in response to the Kerr
field, which is known as cross Kerr nonlinearity. The group
indices for the three different cases can now be calculated as
\begin{eqnarray}
N_g=1+2\pi
Re[\chi]+2\pi\nu_{ac}Re[\frac{\partial\chi}{\partial\delta}],
\end{eqnarray}
\begin{eqnarray}
N^{(0)}_g=1+2\pi
Re[\chi^{(0)}]+2\pi\nu_{ac}Re[\frac{\partial\chi^{(0)}}{
\partial\delta}],
\end{eqnarray}
\begin{eqnarray}
N^k_g&=&1+2\pi Re[\chi^k]+2\pi\nu_{ac}Re[\frac{\partial\chi^k}{%
\partial\delta}]\nonumber\\&&=N^{0}_g+N_2\Omega^2_k,
\end{eqnarray}
where $N_2=\frac{1}{2}\frac{\partial^2N_g}{\partial\Omega^2_k}$ is
the second order refractive index associated with the Kerr field,
while $N_g^0$ is group index when the Kerr field is absent in the
system. The expression for $N_g$ is worked out using traditional
way when the Kerr field is there, but with no contribution from
the higher order nonlinear terms. In principle, there is
contribution from the higher order terms. However, we make this
approach to compare it with $N^k_g$, where the contribution from
the Kerr field enters the susceptibility through our defined
algebra. This then corresponds to the a scientifically reasonable
parameter as compared to the former case. Consequently, for each
case and without showing the indices, the group velocity is read
as
\begin{eqnarray}
v_g=\frac{c}{1+2\pi
Re[\chi]+2\pi\nu_{ac}Re[\frac{\partial\chi}{\partial\delta}]},
\end{eqnarray}
while the advanced time can be estimated from
$\tau_d=\frac{L(N_g-1)}{c}$, correspondingly, where $L$ is the
length of the dispersive medium.

A monochromatic wave-field as a function of angular frequency
$\omega$, position $z$ and time $t$ is written as
$E(z,t)=\frac{1}{2}(E_0e^{i(k(\omega)z-\omega t)}+c.c).$ The
phase, $\Phi$, of the field is also written as
$\Phi=k(\omega)z-\omega t$, which we assume constant during the
propagation time through the medium of length L. Therefore,
$\partial\Phi/\partial t=0$ and $\partial\Phi/\partial\omega=0.$
while $k(\omega)\partial z/\partial t-\omega=0.$
Further,$N_g=n_r(\omega)+\omega\frac{\partial
n_r(\omega)}{\partial\omega}$, while the group velocity dispersion
$D_v$ is written as
$D_v=\frac{\partial}{\partial\omega}(v^{-1}_g)=\frac{1}{c}\frac{\partial
N_g}{\partial\omega}$. The complex wave-number $k(\omega)$ can be
expanded via Taylor series in terms of group index as
\begin{eqnarray}
k(\omega)=\frac{N^{(0)}_g\omega}{c}+\frac{1}{2}(\omega-\omega_0)^2\frac{1}{c}\frac{\partial
N_g}{ \partial\omega}|_{\omega=\omega_0}....
\end{eqnarray}
Incidently, the transit time of a pulse through a material medium
is defined as $T=N_g L/c=k_1 L$, while the spread of transit time
takes the form of $\Delta T=(L\partial N_g/c\partial\omega)\Delta
w,$ where $\Delta w$ appears for the frequency bandwidth. No
significant distortion of the pulse is there if it satisfies the
condition $\Delta T<\tau _{0},$ where $\tau _{0}$ is the
characteristic pulse width. However, if the condition $\partial
N_g/\partial\omega=0$ is satisfied, then there is a negligible
spread in the transit-time-distortion. Successful experiments with
slow and fast light are due to negligible distortion in the output
pulse when $\partial N_g/\partial\omega=0$ while having a larger
group index $N_g$. Obviously, if a system satisfies the condition
for higher order terms of $k(\omega)$, i.e. $k_2,k_3,k_4...=0$,
then there is no distortion in the system at all. The pulse shape
remains unchanged, while the advance time
$\tau_a=\frac{L(N_g-1)}{c}$ and the phase shift are changed. The
refractive index, $n_{r}$, of the medium, $\partial
n_r/\partial\omega$ represents dispersion, while $\partial
N_g/\partial\omega$ gives the distortion of the output signal.
Furthermore, the real part of the higher order terms
$k_{j}(\omega)$ [where $j=1,2,3,.....$] defines the dispersion and
phase distortion, while the imaginary part of $k_{j}(\omega)$
represents the gain [absorption] and the amplitude distortion.
Obviously, the information about the output pulse
$S_{out}(\omega)$ can be extracted from the input pulse
$S_{in}(\omega)$ via the transfer function $H(\omega)$ of the
dispersive medium, and is formulated as
$S_{out}(\omega)=H(\omega)S_{in}(\omega)$, where
$H(\omega)=e^{-ik(\omega)L}$. We choose the Gaussian input pulse
of the form $S_{in}(t)=\exp[-t^2/\tau^2_0]\exp[i(\omega_0+\xi)t]$,
where $\xi$ is the upshifted frequency from empty cavity. The
Fourier transforms of this function is then given by
$S_{in}(\omega)=\frac{1}{\sqrt{2\pi}}\int^{\infty}_{-\infty}S_{in}(t)e^{i\omega
t}dt$. The above integral is worked out as $S_{in}\left( \omega
\right)=\frac{\tau _{0}}{\sqrt{2}}\exp \left[ -\left( \omega
-\omega _{0}-\xi \right) ^{2}\tau^2_0 /4\right]$. By virtue of the
convolution theorem, the output $S_{out}(t)$ can be written  as
$S_{out}(t)=\frac{1}{\sqrt{2\pi}}\int^{\infty}_{-\infty}S_{int}(\omega)H(\omega)e^{i\omega
t}d\omega.$ Integrating the function analytically, we get
\begin{eqnarray}
S_{out}(t)&=&\frac{\tau_0\sqrt{c}}{\sqrt{2i
n_1L+c\tau^2_0}}[1-\frac{i(6\eta_1\eta_2+\eta^3_2)n_2
L}{48c\eta^3_1}]\nonumber\\&&\times\exp[-\frac{\xi^2\tau^2_0}{4}+i(t-\frac{n_0
L}{c})\omega_0]\nonumber\\&&\times\exp[\eta_2+\frac{c\eta^2_2}{(2i
n_1L+c\tau^2_0)}],
\end{eqnarray}
where$\eta_1=\frac{2i n_1L+c\tau^2_0}{4c}$
$\eta_2=\frac{c\xi\tau^2_0-2i n_0L+2ict}{2c}$,  $n_1=\partial
N_g/\partial\omega$ and $n_2=\partial^2 N_g/\partial\omega^2$.
Supposing $n_2=0$. Choosing $\nu_{ac}=1000\gamma$, the gain
doublet occurs at $\delta=30\gamma$.
Here,$\omega/2\pi=\nu_{ac}\pm\delta$, and $\omega=2\pi\nu_p$ is
the probe angular frequency. Furthermore, if the central frequency
of the probe field between the gain doublet is $\nu_c$, then
$\nu_c=(1000\pm30)\gamma$ at $\delta=30\gamma$. The central
angular frequency is $\omega_0=2\pi\nu_c$. Therefore, our
estimation leads to $n_0=-6.16247\times10^5,$ and
$n_{0,k}=-2.43\times10^7$ and their corresponding dispersions
appear as $n_1=\pm 3. 34\times10^8$ and $n_{2}=\pm2.99\times10^7$,
while $n_{1,k}=\pm3.04\times10^8$ and $n_{2,k}=\pm8.43\times10^7$.
The pulse width is given by $\tau_0=3.5\mu s $, which is less than
$t_d$. Consequently, it corresponds to a negligible distortion
meaning lossless characteristics at the anomalous regions in our
atomic system.

Prior to interpreting our results we define necessary terminology.
The field coupled with energy levels $\left|c\right\rangle$ and
$\left|b\right\rangle$ is called a control field when there is no
Kerr effect in the system. In this case we do not include the
additional nonlinear term in the dynamics of the system. This is
done deliberately for a sake of comparison, as there is always
Kerr effect in the system if the coupling laser is intensive. In
this system, the scientifically reasonable results are associated
with analysis of the an additional nonlinear term contributed by
the Kerr effect for the strengthened field. The distinction is
made for the usefulness of the Kerr effect in the enhancement of
superluminal group velocity, a reasonable scientific approach. The
N-type four-level atomic system is reduced to the famous scheme of
Wang \emph{et al.} when the Kerr field is set to zero. In this
approximation, the results of our system agree with the said
reference [see Fig. 2]. Next, in this approximated scheme if we
further exclude one of two coherently driven pump fields, then the
atomic system reduces to a single Raman gain scheme, where only
one gain peak exists. Furthermore, in this approximated setup if
we include the Kerr field with the fourth level, then this forms
the N-type scheme with only one coherent pump field and exhibits a
similar behavior, like the one discussed in Ref. \cite{TNGS2007}.

\begin{figure}[t]
\centering
\includegraphics[width=3in]{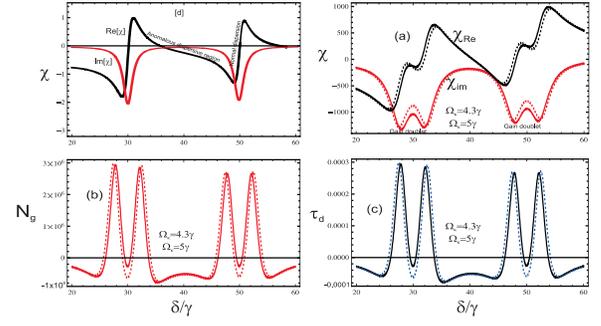}
\caption{Dispersion, gain, group index and advance time without
Kerr effect versus $\frac{\protect\delta}{\protect\gamma}$, such that $c=3\times10^{10}cm/s$, $L=3cm$, $%
\protect\gamma=1MHz$, $\gamma_{ad}=\gamma_{da}=\gamma_{ac}=\gamma_{ca}=%
\gamma_{ab}=\gamma_{dc}=\gamma_{bd}=2.01\protect\gamma$, $\Delta_k=0\protect%
\gamma$, $\protect\lambda=586.9nm$, $\protect\nu_{ac}=10^3\protect\gamma$, $%
\Delta_1=30\protect\gamma$, $\Delta_2=50\protect\gamma$, $\Omega_1=2.5%
\protect\gamma$, $\Omega_2=4\protect\gamma$,
$\Omega_k=4.3\protect\gamma$(solid line) $\Omega_k=5
\protect\gamma$(dashed line).In this case the field is present but
no kerr effect is considered; [d] The reproduction of the Wang
\emph{et al.} results from our analytical results of the
susceptibility under the approximations
$\Omega_k=\gamma_{ad}=\gamma_{da}=\gamma_{bd}=\gamma_{ab}=0$ while
the other parameters are same.}\label{figure1}
\end{figure}
\begin{figure}[t]
\centering
\includegraphics[width=3in]{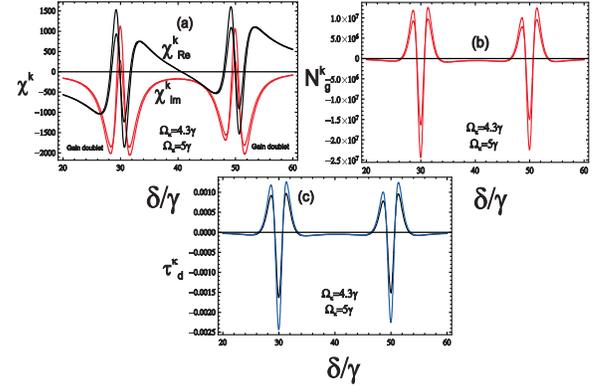}
\caption{Dispersion, gain, group index and advance time for Kerr
effect (the superscript $k$ stands for the Kerr effect) versus
$\frac{\protect\delta}{\protect\gamma}$, when $c=3\times10^{10}cm/s$, $L=3cm$ and $%
\protect\gamma=1MHz$, $\gamma_{ad}=\gamma_{da}=\gamma_{ac}=\gamma_{ca}=%
\gamma_{ab}=\gamma_{dc}=\gamma_{bd}=2.01\protect\gamma$, $\Delta_k=0\protect%
\gamma$, $\protect\lambda=586.9nm$, $\protect\nu_{ac}=10^3\protect\gamma$, $%
\Delta_1=30\protect\gamma$, $\Delta_2=50\protect\gamma$, $\Omega_1=2.5%
\protect\gamma$, $\Omega_2=4\protect\gamma$,
$\Omega_k=4.3\protect\gamma$ (solid line); $ \Omega_k=5
\protect\gamma$(dashed line).Here kerr effect is maximum
contribution.} \label{figure1}
\end{figure}
\begin{figure}[t]
\centering
\includegraphics[width=3in]{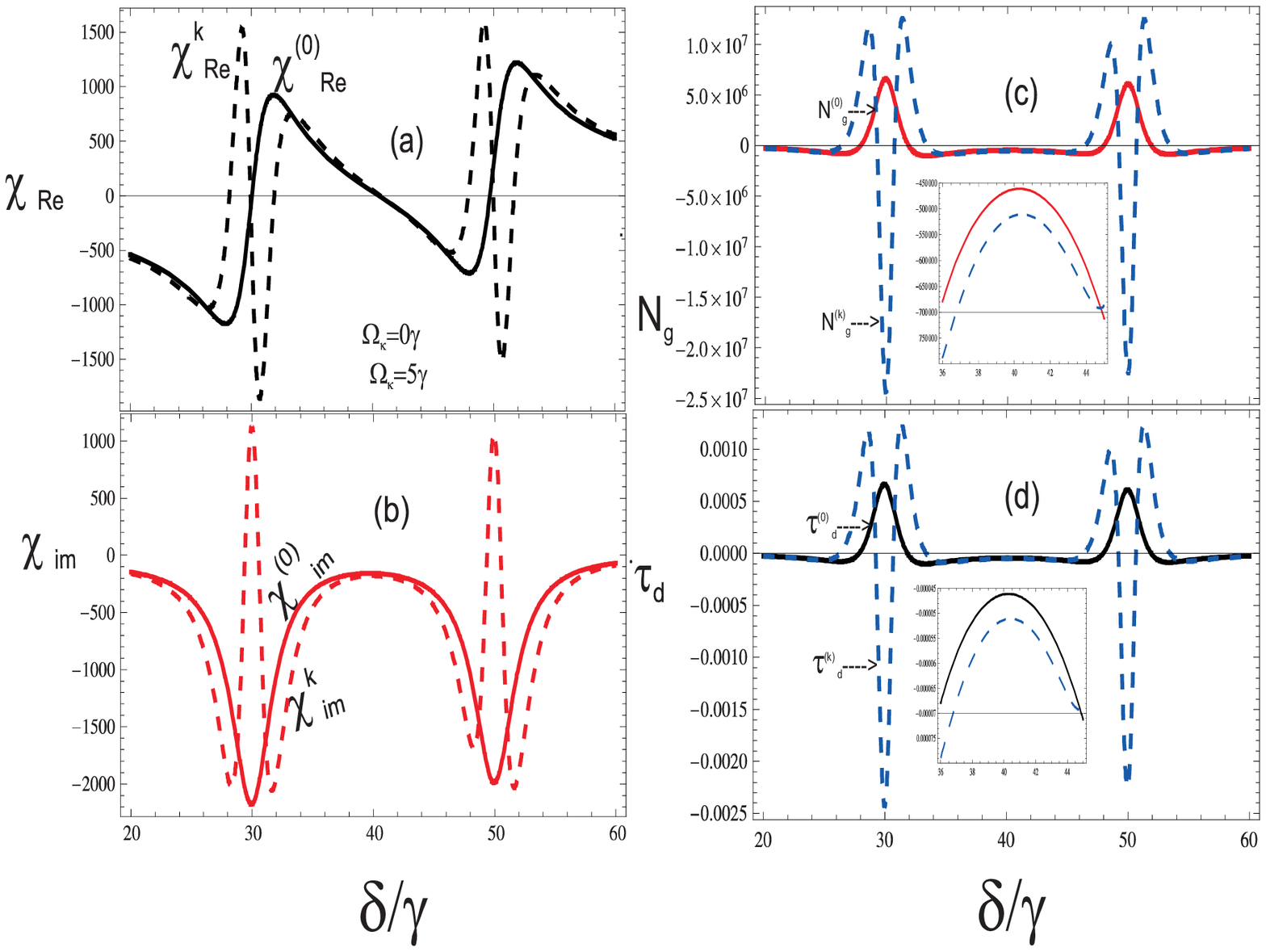}
\caption{Dispersion, gain, group index and advance time for both
the Kerr and without the Kerr field versus
$\frac{\protect\delta}{\protect\gamma}$, when $c=3\times10^{10}cm/s$%
, $L=3cm$ and
$\protect\gamma=1MHz$, $\gamma_{ad}=\gamma_{da}=\gamma_{ac}=\gamma_{ca}=%
\gamma_{ab}=\gamma_{dc}=\gamma_{bd}=2.01\protect\gamma$, $\Delta_k=0\protect%
\gamma$, $\protect\lambda=586.9nm$,
$\protect\nu_{ac}=10^3\protect\gamma, $
$ \Delta_1=30\protect\gamma,$ $\Delta_2=50\protect\gamma$, $\Omega_1=2.5%
\protect\gamma$, $\Omega_2=4\protect\gamma$,
$\Omega_k=0\protect\gamma$ (solid line),
$\Omega_k=5\protect\gamma$ (dashed line).} \label{figure1}
\end{figure}
\begin{figure}[t]
\centering
\includegraphics[width=3in]{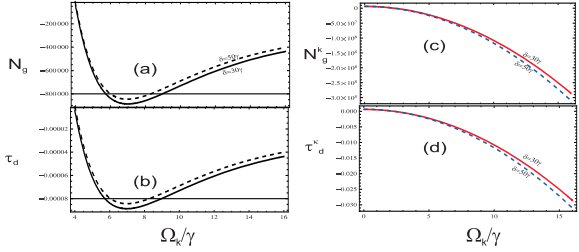}
\caption{The group Index and advance time for both the Kerr and
without the Kerr effect versus $\frac{\Omega_k}{\protect\gamma}$,
when $c=3\times10^{10}cm/s$, $L=3cm$
and $\protect\gamma=1MHz$, $\gamma_{ad}=\gamma_{da}=\gamma_{ac}=%
\gamma_{ca}=\gamma_{ab}=\gamma_{dc}=\gamma_{bd}=2.01\protect\gamma$, $%
\Delta_k=0\protect\gamma$, $\protect\lambda=586.9nm$, $\protect\nu_{ac}=10^3%
\protect\gamma$, $\Delta_1=30\protect\gamma$, $\Delta_2=50\protect\gamma$, $%
\Omega_1=2.5\protect\gamma$, $\Omega_2=4\protect\gamma$,
$\protect\delta=30\gamma$ (solid line); $\delta=50\protect\gamma$
(dashed line).  } \label{figure1}
\end{figure}
\begin{figure}[t]
\centering
\includegraphics[width=3in]{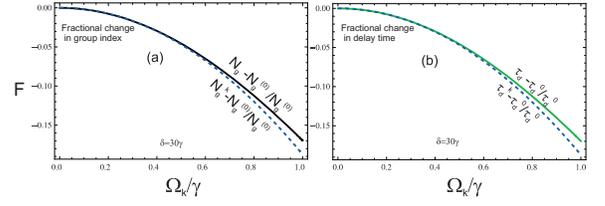}
\caption{The fractional change in the group index and the advance
time for both the Kerr and without the Kerr effect versus
$\frac{\Omega_k}{ \protect\gamma}$, when $\protect\gamma=1MHz$,
$\gamma_{ad}=\gamma_{da}=
\gamma_{ac}=\gamma_{ca}=\gamma_{ab}=\gamma_{dc}=\gamma_{bd}=2.01\protect
\gamma$ , $\Delta_k=0\protect\gamma$, $\protect\lambda=586.9nm$,
$\protect \nu_{ac}=10^3\protect\gamma$,
,$\Delta_1=30\protect\gamma$, $\Delta_2=50 \protect\gamma$,
$\Omega_1=2.5\protect\gamma$, $\Omega_2=4\protect\gamma$, $
\protect\delta=30\protect\gamma$, $c=3\times10^{10}cm/s$, $L=3cm$,
for $N_g-N^{(0)}/N^{(0)}$ (solid line); and for
$N^k_g-N^{(0)}/N^{(0)}$ (dashed line) same solid and dashed line
for time fraction.} \label{figure1}
\end{figure}
\begin{figure}[t]
\centering
\includegraphics[width=3in]{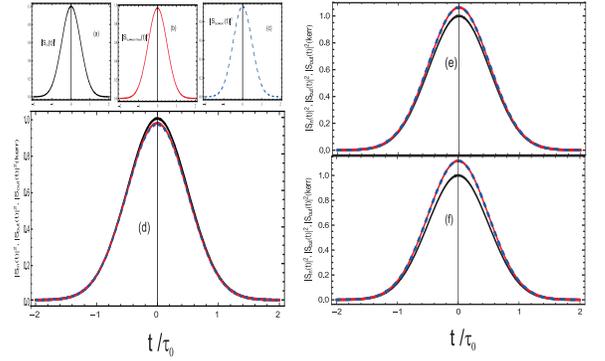}
\caption{Input and output pulse intensity [a] input (black) [b]
output without the Kerr effect (red) [c] with the Kerr effect
(dashed blue) [d ,e,f]  for $\protect\gamma=1MHz$,
$\gamma_{ad}=\gamma_{da}=\gamma_{ac}=\gamma_{ca}=
\gamma_{ab}=\gamma_{dc}=\gamma_{bd}=2.01\protect\gamma$, $\Delta_k=0\protect%
\gamma$, $\protect\lambda=586.9nm$,
$\protect\nu_{ac}=10^3\protect\gamma$,
$\Delta_1=30\protect\gamma$, $\Delta_2=50\protect\gamma$, $\Omega_1=2.5%
\protect\gamma$, $\Omega_2=4\protect\gamma$,
$\Omega_3=4.3\protect\gamma$, $L=0.03m$, $c=3\times10^{8}m/s$,
$\omega_0=2\pi\times(1000\pm30)\gamma\times rad$, $\tau_0=3.5\mu
s$, [d]$\xi=0.001\gamma\times rad$[e] $\xi=0.005\gamma\times
rad$[f]$\tau_0=10.5\mu s$, $\xi=0.001\gamma\times rad$}
\label{figure1}
\end{figure}
The cross Kerr nonlinearity is deliberately introduced in the
medium with a goal to induce maximum coherence via a large Kerr
effect. The susceptibility displays two pairs of double-peak gain
lines, a novel and interesting behavior of the system. The physics
here is different than earlier approaches
\cite{PRAoptimal,eurolett,opticslett,opticslett01,PRA-instable}].
The coupling of an additional Kerr field splits the ground energy
level to a doublet. Subsequently, the photons added to the probe
field by the two pump fields follow two different paths provided
by the two dressed states. In this way one pair of the gain
appears for each coherently driven pump field. The additional Kerr
field modifies the gain feature, but the created anomalous regions
remain almost lossless similar to the Ref. \cite{Wang2000}.
Therefore, the induced Kerr effect which drastically increases the
group index is still practical. The anomalous regions at the three
position of dispersion is controllable with strength of the Kerr
field. Intriguingly, these controllable regions may resolve the
issues of the distortion caused by incompatibility of the width of
probing pulse and the anomalous regions for the transparency. In
Fig. 2 and Fig. 3 we show the plots of dispersion and gain
spectrum under similar parameters for the Kerr free system
($\chi$) as well as for a Kerr effect ($\chi^k$) for condition of
two photon resonances: $(\delta=\Delta_1$ or $ \delta=\Delta_2).$
The slopes of the dispersions between the gain regions are
anomalous for both $\chi$ and $\chi^k$ but steeped for the Kerr
one. Correspondingly, negative group index for the Kerr effect is
larger than that of the Kerr free case. This then corresponds to
an enhancement in the group velocity of the superluminal light
pulse for the Kerr effect, meaning advancement in time under the
same set of parameters [Fig. 2(c) and Fig. 3(c)] respectively.
Next, we display the behavior of the dispersion, gain, and their
corresponding group index, and the advanced time with the probe
field detuning when the Kerr field vanishes. Obviously, there are
two gain peaks in the spectrum at the two photon resonance points:
$ \delta=\Delta_1$ or $\delta=\Delta_2$. This behavior is in
accord with the scheme of Wang \emph{et al.} Consequently, the
slope of the dispersion becomes steeply anomalous in these
resonance points. The group index and the negative time delay are
$-2.5\times10^7$ and $-2.5ms$, respectively [see Fig. 4(a-d) dash
lines]. These time delays are always greater than the negative
delay time of the Wang et al. experiment for both the Kerr effect
and the Kerr free systems. Further, the group indices in between
the gain lines are also negative but have enhanced profile for the
Kerr field [see the inset Fig, 4(c-d)]. The group index without
the Kerr field at the point $ \delta=40\gamma$ is $-500000$, and
in the presence of Kerr effect is $-550000$. The time delays are
then $-50\mu s$ and $-55\mu s$ for Kerr effect and without the
Kerr effect, respectively. The advance time of the pulse is
increased by $5\mu s$ with the Kerr effect. The group index varies
with $\Omega_k/\gamma$ for $ \delta=30\gamma$ or
$\delta=50\gamma$, and is always more negative for Kerr effect
[see Fig. 5]. Group index in the presence of the Kerr effect for
$\Omega_3=16\gamma$ is $N^k_g=-3.10046\times10^8$, and in the
absent of the Kerr effect it is $N_g=-3.47245\times10^5$ while
having $\delta=30\gamma$ [see Fig. 5(a-d)]. The group velocity for
a Kerr free system is $v_g=c/N_g$, while it is $v^k_g=c/N^k_g$
when there is no Kerr effect in the system. The corresponding
group velocities are $-68611.4cm/s$ and $-96.7598cm/s$ under
similar parameters while having $\delta=30\gamma$ and
$\Omega_3=16\gamma$, respectively. Moreover, the negative group
delay time is $-48\mu s$ when there is no Kerr effect in the
system. However, in the presence of the nonlinear Kerr effect the
negative time delay is $-30 m s$ while having $\delta=30\gamma$
and $\Omega_3=16\gamma$, respectively. Furthermore, the comparison
of fractional change in the group index and in the advance time
also confirms the dominant stable behavior for the Kerr effect in
the system. The negative group index means sooner arrival relative
to the velocity of the pulse in vacuum. The pulse travelling in a
medium with a larger negative group index, in fact, arrives much
sooner than pulse seeing the medium with relatively lesser
negative group index. Consequently, the superluminal Gaussian
pulse leaves the medium for the Kerr effect sooner by $30 m s$
than the pulse leaving the medium when there is no Kerr effect in
the system.

In Ref. \cite{Akulshin01} the incompatibility of power broadening
of the drive bi-chromatic laser field with closely spaced ground
hyperfine splitting resulted in instability in comparison with a
situation when a monochromatic field was used as mentioned by
Akulshin and his coworker in another Ref. \cite{PRA-instable} .
Furthermore, the group index with a Kerr effect has a directly
proportionality with the drive field intensity presented for
analysis of the experimental result while the instability limit
appears from a Kerr free result for group index [see Ref.
\cite{Akulshin01}]. In principle, if the drive field intensity is
large enough then a coherence effect must be induced by a Kerr
effect as presented in this manuscript in comparison to our
assumed Kerr-free system. Consequently the induced coherence
effect overcome the fixed incoherence power broadening of the
bi-chromatic field. Correspondingly, if we do not include the Kerr
effect in our estimation then the system suffers from the
instability limit which is in fact scientifically not reasonable
as shown in Fig. (5) [compare the Kerr-free and Kerr effect
cases]. In addition, the incoherence mechanism may be more reduced
if a medium with very slow ground state relaxation is considered.
For example, a cell with buffer gas [18] or a cell with an
anti-relaxation coating [19] may facilitates an experiment to
overcome ground state relaxation.

Conceptually, enhancement in negative group index is not a
necessary condition for a useful superluminal effect rather it
requires undistort retrieved pulse, in addition.  Earlier,
physical interpretation of superluminality by virtual reshaping is
now not always reasonable \cite{Wangpra,book1,book2} with the
explanation of amplification of the front edge with the relatively
absorbtion of its tail. Here, in the pulse distortion measurement
we kept the probe pulse bandwidth much smaller than gain lines
separation and interaction time to avoid resonances with the Raman
transitions frequencies for the probe pulse. Consequently, there
is no amplification of the front edge of the pulse. This means
that both the front edge and the tail would be amplified if the
atoms of the medium amplify the probe pulse. However, this is not
in accord with the earlier claims. Obviously, the superluminal
light propagation arises due to the anomalous dispersion regions
created in between the nearby Raman gain resonances of our system.
In fact, if the gain becomes large, its effect appears as a
compression of the pulse \cite{Wangpra} [see Fig. 7]. The detail
analysis reveals that the pulse distortion measurement fully
agrees with the Wang \emph{et al.} studies even with the second
order perturbation limit. Evidently, in our system it is shown
that pulse shape remains preserve over the very small region in
between the gain lines for some specific value of the upshift
frequency of a little change in the probe pulse width. The
shifting of this upshift frequency (the probe field pulse width)
results in compression (enhancement) of the retrieved Gaussian
pulse. Unlikely, this behavior does not agree with the earlier
interpretation and is more likely agree with Wang \emph{et al},
presented results and interpretation. Satisfactorily, we are
providing analytical results to the literature where this
transparency behavior may be demonstrated in a laboratory and
proceeded beyond the existing literature.

In conclusion, we induce quantum coherence effect in a
gain-assisted N-type 4-level atomic system via nonlinear Kerr
effect by the coupled intense monochromatic laser field. The
coherence avoids the instability limit constrained by the earlier
traditional approach to superluminality. The proposed scheme
displays novel, amazing and useful behavior of two-pair gain lines
processes. Consequently, a remarkable enhancement in the
superluminal effect on the Gaussian pulse is seen. Quantitatively,
under experimentally feasible parameters, the pulse advances by
almost $30 ms$ more than that of the Kerr free system of the Wang
\emph{et al.} The proposed system is lossless like Wang \emph{et
al} but with the advantages of \emph{multiple}
\textbf{controllable} anomalous regions, \emph{significantly
enhanced} and \textbf{stable} superluminal behavior, and
\textbf{relaxed} temperature, states of matter along with their
isotropy, or anisotropy conditions, respectively. The control of
multiple anomalous regions may insure an undistorted retrieved
pulse due to their compatibility with the probe pulse width. The
superluminal effect may also be more enhanced if inhomogeneous
solid state medium with a large enough atomic density is selected.
Incorporating these aspects in an experiment may modify some
related current experimental technologies to help improve applied
aspects of a superluminal light pulse in a medium.

\section{Appendix}
The parameters $\beta_{1,2}$ in Eq. (4) and the parameters
$\beta_3$, $T_{1,2}$, $T_3$, $K_{1,2}$ in Eq. (7) are listed
bellow:
\begin{eqnarray}
\beta_{1,2}=\frac{4|\Omega_{1,2}|^2[P_{1,2}+ L_{1,2}]}
{4(\gamma_{ac}-i\delta)(\gamma_{ab}-i(\delta-\Delta_k))+\Omega^2_k},
\end{eqnarray}
where
\begin{eqnarray}
P_{1,2}=\frac{2\gamma_{ad}(\gamma_{ab}-i(\delta-\Delta_k))}{(\gamma_{ca}+\gamma_{da})(\gamma^2_{ad}+\Delta^2_{1,2})},
\end{eqnarray}
and
\begin{eqnarray}
L_{1,2}=\frac{4(\delta-\Delta_k+i\gamma_{ab})(\delta-\Delta_{1,2}-\Delta_k+i\gamma_{bd})+\Omega^2_k}{(\gamma_{ad}+i\Delta_{1,2})R_{1,2}},
\end{eqnarray}
with
\begin{eqnarray}
R_{1,2}=[4(\Delta_{1,2}-\delta-i\gamma_{dc})(\Delta_{1,2}+\Delta_k-\delta-i\gamma_{bd})-\Omega^2_k].
\end{eqnarray}
while
\begin{eqnarray}
\beta_3=\frac{2\gamma_{ad}[(\gamma^2_{ad}+\Delta^2_2)\Omega^2_1+(\gamma^2_{ad}+\Delta^2_1)\Omega^2_2]}{(\gamma_{ca}+\gamma_{da})(\gamma_{ac}-i\delta)
(\gamma^2_{ad}+\Delta^2_1)(\gamma^2_{ad}+\Delta^2_2)},
\end{eqnarray}

\begin{eqnarray}
T_{1,2}=\frac{|\Omega_{1,2}|^2}{(\gamma_{ac}-i\delta)(\delta-\Delta_{1,2}+i\gamma_{dc})(\Delta_{1,2}-i\gamma_{ad})}
\end{eqnarray}

\begin{eqnarray}
T_3=(\gamma_{ac}-i\delta)(\gamma_{ab}-i(\delta-\Delta_k)),
\end{eqnarray}
and
\begin{eqnarray}
K_{1,2}=\frac{[\frac{i}{\Delta_{1,2}+\Delta_3-\delta-i\gamma_{bd}}-\frac{\alpha_{1,2}+x_{1,2}}{x_{1,2}\gamma_{ac}-ix_{1,2}\delta}]|\Omega_{1,2}|^2}{2\alpha_{1,2}},
\end{eqnarray}
while
\begin{eqnarray}
\alpha_{1,2}=T_3(\gamma_{ad}+i\Delta_{1,2})(\gamma_{dc}-i(\delta-\Delta_{1,2})),
\end{eqnarray}
and
\begin{eqnarray}
x_{1,2}&=&[\gamma_{dc}-i(\delta-\Delta_{1,2})]^2(\gamma_{ad}+i\Delta_{1,2})\nonumber\\&&\times[\gamma_{bd}-i(\delta-\Delta_{1,2}-\Delta_k)].
\end{eqnarray}

\end{document}